\documentclass{openjournal}

\usepackage{xcolor}
\usepackage{textgreek}
\usepackage[utf8]{inputenc}
\usepackage[english]{babel}
\usepackage{amsmath}
\let\oldAA\AA
\renewcommand{\AA}{\text{\normalfont\oldAA}}
\usepackage{subfigure}
\usepackage{booktabs}

\usepackage{hyperref}
\hypersetup{
    unicode, 
    colorlinks=true,
    linkcolor=linkcolor,
    citecolor=linkcolor,
    filecolor=linkcolor,
    urlcolor=linkcolor,
}
\usepackage{color,colortbl}
\definecolor{linkcolor}{rgb}{0.0,0.3,0.5}
\usepackage{tensind}
\tensordelimiter{?}
\DeclareGraphicsExtensions{.bmp,.png,.jpg,.pdf}
\usepackage{verbatim}
\usepackage[normalem]{ulem}
\usepackage{orcidlink}
\usepackage{soul}

\graphicspath{ {./figs/} }

\begin{document}
\title{Vision-Based CNN Prediction of Sunspot Numbers from SDO/HMI Images}

\author{Fabian C. Quintero-Pareja$^{a}$\orcidlink{0009-0001-1851-059X}, Diederik A. Montano-Burbano$^{a}$\orcidlink{0009-0006-7441-6342}, Santiago Quintero-Pareja$^{a}$\orcidlink{0009-0006-2434-8273}, and D. Sierra-Porta$^{a,*}$\orcidlink{0000-0003-3461-1347}}
\email[Fabian C. Quintero-Pareja: ]{parejaf@utb.edu.co}
\email[Diederik A. Montano-Burbano: ]{dmontano@utb.edu.co}
\email[Santiago Quintero-Pareja: ]{squintero@utb.edu.co}
\email[D. Sierra-Porta: ]{dporta@utb.edu.co}
\affiliation{Universidad Tecnológica de Bolívar. Escuela de Transformación Digital. Parque Industrial y Tecnológico Carlos Vélez Pombo Km 1 Vía Turbaco. Cartagena de Indias, 130010, Colombia}
\affiliation{$^{*}$ Corresponding author: dporta@utb.edu.co (D. Sierra-Porta)}

\begin{abstract}
    Sunspot numbers constitute the longest and most widely used record of solar activity, with direct implications for heliophysical research and space-weather applications. Traditional sunspot counting relies on visual inspection and algorithmic feature-detection pipelines, which can be affected by observer-dependent choices, image quality, and methodological variability across implementations. Recent advances in deep learning, particularly convolutional neural networks (CNNs), enable the direct use of solar imagery for automated image-to-scalar regression, reducing the need for explicit, handcrafted feature design. In this work, we present a supervised vision-based framework to estimate the daily sunspot number from full-disk continuum images acquired by the Helioseismic and Magnetic Imager (HMI) onboard NASA’s Solar Dynamics Observatory (SDO). Images from 2011–2024 were paired with daily sunspot numbers from the SILSO Version 2.0 dataset maintained by the Royal Observatory of Belgium. After preprocessing and augmentation, a CNN was trained to infer a scalar sunspot number directly from pixel data at the observation time of each image. The proposed model achieved strong performance on an independent test split ($R^2$=0.964, RMSE=9.75, MAE=6.74), indicating close agreement with SILSO reference values across a broad activity range. In comparison with prior studies, we position this approach as a competitive and conceptually simple alternative for direct image-based estimation, complementing time-series forecasting models that target monthly means or smoothed indices. Interpretability analyses using Grad-CAM and Integrated Gradients indicate that the network consistently attributes relevance to sunspot-bearing regions when forming its estimates. These results highlight the potential of deep vision-based approaches for scalable solar monitoring and automated estimation of classical heliophysical indices. Future work should explore multimodal fusion with additional observables (e.g., magnetograms) and standardized cross-cycle benchmarks to strengthen robustness under changing solar conditions.
\end{abstract}



\section{Introduction}
\label{sec:intro}

Sunspots represent a fundamental and enduring measure of solar activity, with sunspot numbers providing the longest continuous record of solar variability, spanning more than 400 years \citep{2022SoPh..297..158H, 2024AJ....167...52Z}. The sunspot number is one of the most widely used indices in solar-terrestrial research due to its strong correlations with multiple space weather phenomena \citep{2012EPJP..127...43C}. For example, the frequency and energy of solar flares, as well as the rate of coronal mass ejections, are well correlated with sunspot numbers, while cosmic ray fluxes exhibit an anticorrelated pattern over the solar cycle \citep{2012EPJP..127...43C, 2002A&A...386..313O}.

Sunspot observation is one of the oldest continuous scientific endeavors, with telescopic data spanning centuries and compiled into well-calibrated time series that reflect solar magnetic activity \citep{2022SoPh..297..158H, 2025RASTI...4...24M, 2024SoPh..299..156H, 2024AJ....167...52Z}. The foundation of modern sunspot quantification lies in the relative sunspot number, defined as $k(10g + f)$, where $f$ represents the number of individual sunspots, $g$ the number of sunspot groups, and $k$ a correction factor accounting for observer and instrumental differences.

{Traditional manual methods for sunspot detection suffer from human subjectivity and inconsistencies, while automated techniques based on spatial filtering may compromise image resolution or distort physical sunspot properties \citep{sarsembayeva2021detecting, sarsembayeva2025solar, 1990ApJ...356..733B}. Measurement accuracy has been shown to depend heavily on both the method and image quality, with systematic discrepancies noted between photographic records and sunspot drawing archives \citep{2001MNRAS.323..223B}. 
Importantly, algorithmic detection pipelines also involve design choices (e.g., thresholds, morphology, and grouping rules) that can affect outputs; likewise, neural-network approaches depend on architectural and hyperparameter selections. In this work, the primary advantage of CNN-based regression is not the absence of methodological choices, but the ability to learn task-relevant representations directly from data in a consistent, end-to-end manner once the training protocol is fixed.}

Technological advances have facilitated the development of high-resolution observational tools such as CCD imaging, spectroscopy, satellite platforms, and radio diagnostics, alongside tools for automated sunspot group area calculation, {including software like S.C.A.T. (Solar Cycle Analyzer Tool; \citealp{cedazo2020improving}).} Recent efforts have leveraged image processing and deep learning techniques to improve automatic sunspot detection, although accurate grouping of sunspots remains a challenge \citep{2024AJ....167...52Z}. Modern computer vision methods, including Canny edge detection and topological analysis, have been applied to solar images from NASA’s Solar Dynamics Observatory (SDO) for feature identification \citep{2024arXiv240502545S, rong2014improved, cheng2012improved}.

Recently, \cite{sierra2024predicting} proposed a machine learning method {to estimate sunspot numbers from solar imagery}, leveraging topological feature extraction prior to modeling. In contrast, our work employs fully automated convolutional neural networks {to directly map full-disk solar images to sunspot number estimates}, removing the need for handcrafted feature design and enabling end-to-end regression learning. {Note that we use the term ``prediction'' in the machine-learning sense of inferring a target value from input data at the same observation time (i.e., an image-to-index estimation), not as a forecast of future sunspot numbers.} The rise of deep learning—especially Convolutional Neural Networks (CNNs)—has revolutionized solar physics by enabling the analysis of massive, high-dimensional datasets. SDO alone generates over 1.5 terabytes of data daily, {motivating systematic, automated processing pipelines before the widespread adoption of deep learning} \citep{love2020analyzing, Martens2012}. {Here, by ``traditional'' we refer to rule-based image processing and classical machine-learning pipelines that rely on manually engineered features; extensive benchmarking efforts for these methods exist in the literature (e.g., FLARECAST) \citep{Georgoulis2021}.} Deep learning models can extract hierarchical spatial features directly from raw observational data without the need for handcrafted input representations \citep{love2020analyzing, chola2022detection, diaz2022towards}.

CNNs have shown outstanding performance in characterizing solar active regions. Recognition accuracies above 95\% have been reported for magnetic classification of sunspot groups, with Alpha-type accuracies reaching 98\% and Beta-types above 88\% \citep{2020FrP.....8...45F, 2019AdAst2019E..27F}.

Solar flare prediction is another key domain where CNNs have outperformed traditional methods. Recent models achieve True Skill Statistic (TSS) scores of $0.749 \pm 0.079$ for $\geq$M-class flares and $0.679 \pm 0.045$ for $\geq$C-class, surpassing earlier techniques \citep{2023aike.conf...83P, 2020ApJ...891...10L}. Some architectures exceed TSS 0.80 for M-class prediction tasks \citep{shen2024deep, xu2025solar}.

Advances in attention-based deep learning have also enabled full-disk solar flare prediction, overcoming the limitations of early models that {focused only on central disk regions ($\pm 30^\circ$ to $\pm 45^\circ$ in heliocentric angle from disk center, i.e., distance from disk center in angular coordinates)} \citep{2023aike.conf...83P, 2018ApJ...856....7H, 2020ApJ...891...10L}. These models have demonstrated average recall rates near 0.52 for near-limb flares beyond $\pm 70^\circ$ \citep{2023aike.conf...83P}.

The field now extends beyond classification, incorporating solar radio spectrum analysis where CNNs and LSTMs are used to identify and classify solar radio bursts across vast datasets \citep{2022Univ....8..656L, 2018A&A...618A.165Z}. Operational systems integrate ensemble learning, physics-informed models, and interpretable AI techniques such as attention maps and gradient attribution to enhance transparency \citep{ji2023interpretable}.

The synergy between deep computer vision and solar physics big data has enabled real-time extraction of features from solar images with remarkable accuracy. Architectures like U-Net and ResUNet++ further refine segmentation performance for complex solar structures \citep{2025arXiv250616194B, jha2019resunet++, ronneberger2015u}.

While deep learning has achieved impressive results in solar physics, several open challenges remain. Chief among these is the scarcity of high-quality, labeled datasets, as creating ground truth annotations demands expert effort and time. This scarcity is particularly acute for rare events such as X-class flares. Another limitation is model generalization: CNNs trained on specific datasets often struggle under changing observational conditions or across solar cycles. Instrumental variation, atmospheric distortion, and temporal evolution introduce domain shifts that hinder cross-cycle robustness.

Computational constraints also arise in real-time applications, where deep models may be too resource-intensive for timely estimation. Additionally, interpretability remains an issue: most CNNs behave as black boxes, limiting the ability of researchers to trust or explain model outputs. Finally, inconsistent evaluation metrics and benchmarking datasets across the literature hinder fair comparison between methods. Most applications to date focus on classification rather than continuous regression tasks, leaving a gap in the direct prediction of quantitative indices such as sunspot numbers or magnetic field strength.

In this study, we present a deep learning framework to estimate the daily sunspot number directly from full-disk digital images acquired by the Helioseismic and Magnetic Imager (HMI) aboard NASA's Solar Dynamics Observatory (SDO). We compile a dataset of continuum solar images from 2011 to 2024 and pair these with corresponding daily sunspot numbers from the SILSO Version 2.0 database maintained by the Royal Observatory of Belgium. Our approach leverages Convolutional Neural Network (CNN) architectures trained in a supervised regression setting, aiming to predict the continuous sunspot number as a function of the visual solar disk appearance alone.

While prior research has extensively applied CNNs to tasks such as solar flare classification, sunspot group type recognition, or magnetic topology identification, few efforts have directly targeted the prediction of the scalar sunspot number from image data. This task poses distinct challenges, particularly due to the complex spatial distribution and varying contrast of sunspots across the solar disk, as well as the absence of public benchmarks for image-to-scalar sunspot regression. By addressing this gap, our work contributes a novel methodology that bridges solar image processing and quantitative sunspot analysis, expanding the scope of deep learning applications in operational space weather monitoring.

Our results show that deep learning techniques are not only capable of modeling visual solar patterns but also hold potential for quantitatively bridging observational solar imagery and fundamental heliophysical indices. This work thus contributes to {expanding the toolkit for space weather monitoring by integrating automated vision-based approaches into the estimation of classical solar metrics.}

\section{Data and Methods}
\label{sec:data_meth}

\subsection{Sunspot Number Data}

Since 1981, the Solar Influences Data Analysis Center (SIDC) has served as the World Data Center for sunspot numbers, operating the SILSO (Sunspot Index and Long-term Solar Observations) project, which maintains the longest-running time series of solar activity measurements. {The international sunspot number provided by SILSO is a vital reference for numerous scientific disciplines and is widely used by the solar--terrestrial research community, including international institutions and organizations.}

The revised SILSO Version 2.0 dataset, released in 2015, incorporates recalibrated sunspot counts with improved consistency across observatories and over time. This dataset continues to rely primarily on visual hand-drawn sunspot counts from a network of ground-based stations, with the pilot station at Specola Solare Ticinese in Locarno, Switzerland, serving as the primary reference \citep{2022SoPh..297..158H}. The Version 2.0 revision corrected historical biases, homogenized the long-term series, and introduced an internally consistent standard that is now widely adopted in solar physics research. 

In this study, we used the daily total sunspot number covering the period from January 1st, 2011 to February 29th, 2024. This interval was selected to align with the availability of continuum solar images from the Helioseismic and Magnetic Imager (HMI) onboard NASA’s Solar Dynamics Observatory (SDO). The daily sunspot number was extracted as a single scalar value representing the global level of solar magnetic activity, which served as the target variable for our supervised regression task. In addition to its long-term stability and scientific acceptance, this dataset is particularly {well suited as a reference target for supervised learning because} it encodes the aggregate visual manifestation of sunspots across the entire solar disk, making it directly comparable to the image-based features learned by convolutional neural networks.

\begin{figure}[htb]
    \centering
    \includegraphics[width=0.7\textwidth]{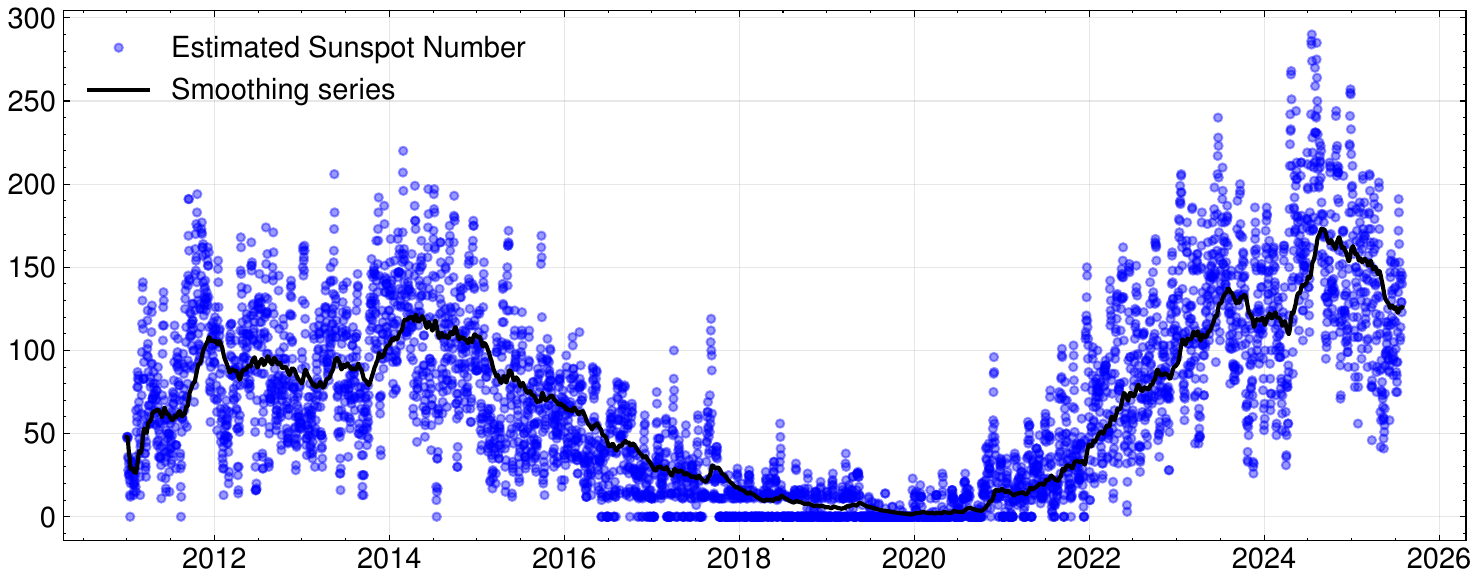}
    \caption{Daily sunspot number time series from SILSO (Version 2.0) between 2011 and 2024, including the smoothed solar cycle trend. This dataset served as the reference target for model training and evaluation.}
    \label{fig:silso_timeseries}
\end{figure}

Figure~\ref{fig:silso_timeseries} shows the daily sunspot number from SILSO (\url{https://www.sidc.be/SILSO/datafiles}) during the selected study period, including its smoothed profile that highlights the underlying 11-year solar cycle. The dataset spans the descending phase of Solar Cycle 24, the minimum around 2019, and the ascending phase of Solar Cycle 25 through 2024. This coverage ensures that the model is trained and evaluated across both low-activity and high-activity regimes, providing a more robust assessment of its generalization capacity.

\subsection{Solar Image Data from SDO/HMI}

The Solar Dynamics Observatory (SDO), launched by NASA in 2010, continuously acquires full-disk images of the Sun across multiple spectral channels. 

{For this study, we employed HMI continuum intensity observations at 6173~$\AA$, accessed as publicly available full-disk quicklook JPEG images from the SOHO/SDO reprocessing archive (\url{https://soho.nascom.nasa.gov/data/REPROCESSING/Completed/}).} This choice provides a standardized and easily accessible image stream suitable for computer-vision ingestion, while preserving the photospheric contrast patterns associated with sunspot morphology. We note, however, that the upstream pipeline used to generate quicklook products (including calibration parameters and rendering choices) may evolve over time; therefore, our use of quicklooks prioritizes practical reproducibility of data access and formatting rather than guaranteeing strict time-invariant photometric calibration. This wavelength samples the photospheric layer of the Sun and provides high-resolution images of the visible solar disk, where sunspots appear as dark regions due to their lower temperature relative to the surrounding photosphere. Continuum images are therefore particularly well suited for studies that require direct correspondence between sunspot morphology and the sunspot number index.

{We selected one HMI continuum image per day from 2011-01-01 to 2024-02-29, choosing the frame closest to 00{:}00~UTC to ensure daily consistency and temporal alignment with the SILSO sunspot index.} Each raw image, originally recorded at 4096$\times$4096 pixels, included metadata margins and peripheral artifacts. To standardize the dataset and prepare it for convolutional neural network (CNN) ingestion, we cropped the images to isolate the solar disk, removed extraneous borders, and downsampled them to 512$\times$512 pixels using bilinear interpolation. {This preprocessing enforces a fixed input geometry for the CNN while retaining the full-disk context.} Pixel intensities were normalized to the range [0,1] to stabilize the training process. In addition, data augmentation techniques such as {random horizontal flips, small rotations (within $\pm 10^{\circ}$), and isotropic rescaling within $\pm 3\%$ were applied during training} to enhance generalization and reduce overfitting. Each processed image was paired with the corresponding daily sunspot number from SILSO as its scalar regression target, thus enabling supervised learning. {The resulting dataset contains 4778 daily images, split into 3822 for training and 956 for testing.}

\begin{figure}[htb]
    \centering
    \includegraphics[width=0.9\textwidth]{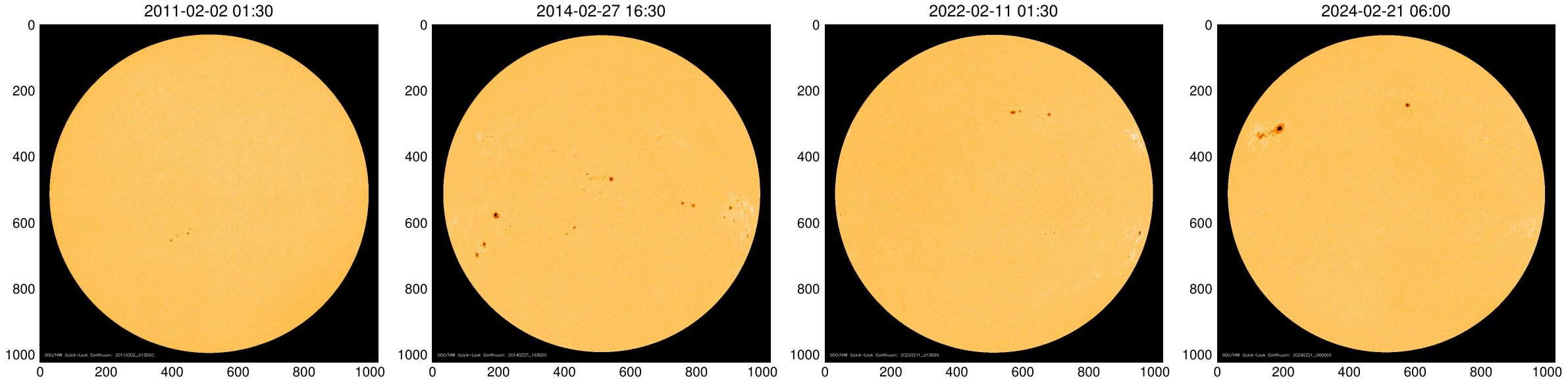}
    \caption{Representative continuum intensity images from the Helioseismic and Magnetic Imager (HMI) aboard SDO (\url{https://soho.nascom.nasa.gov/data/REPROCESSING/Completed/}). The images illustrate solar disk evolution across different phases of solar activity (2011–2024). These full-disk images were preprocessed and paired with SILSO daily sunspot numbers for supervised regression.}
    \label{fig:sdo_examples}
\end{figure}

Figure~\ref{fig:sdo_examples} illustrates representative HMI continuum images across different phases of solar activity, ranging from solar minimum to maximum. These examples highlight the pronounced morphological variability of sunspots on the photosphere, which the CNN model is designed to capture and map to the daily sunspot number. By combining the temporal continuity of the SILSO index with the spatial richness of SDO/HMI imagery, this dataset provides a robust foundation for training deep learning models that bridge visual solar features with quantitative heliophysical indices.

\subsection{Model Architecture and Training Procedure}
{We developed a supervised regression model based on a Convolutional Neural Network (CNN) architecture tailored for image-to-scalar estimation. The final model consists of a hierarchy of convolutional stages using $3\times 3$ kernels and Swish activations \citep{mercioni2020p, 2024arXiv240708232R}. Each stage applies two convolutional layers followed by batch normalization and spatial downsampling via max pooling, progressively increasing the number of feature channels (32, 64, 128, 256, and 512) while reducing spatial resolution. The convolutional backbone is followed by a global average pooling layer and a lightweight fully connected head (Dense 256 with dropout, Dense 128), ending in a single linear neuron to output a continuous sunspot number estimate.}

{The model was trained using the Huber loss function \citep{gupta2020robust}, which combines the robustness of the Mean Absolute Error (MAE) with the sensitivity of the Mean Squared Error (MSE), making it well-suited for handling outliers in solar activity. Optimization was performed using the Adam algorithm with a learning rate of $10^{-4}$ \citep{2019arXiv190409237R}. The dataset comprises 4778 daily image--label pairs and was split chronologically into training (3822 images) and testing (956 images) subsets to preserve temporal consistency and prevent information leakage. Model selection and early stopping were driven by a validation subset carved out from the training period (using a fixed hold-out fraction of the training data), and training was run for up to 100 epochs with early stopping on validation loss to mitigate overfitting.}

This splitting strategy is primarily designed to prevent information leakage in a time-ordered dataset. Nevertheless, a purely chronological split does not guarantee identical activity-level distributions across subsets (e.g., different proportions of rising/declining phases), and a hold-out validation subset drawn from the training period does not fully assess generalization under distribution shift. A more rigorous assessment would require blocked cross-validation, cross-cycle holdouts, or cross-instrument evaluation protocols. We therefore interpret the reported test performance as evidence of strong within-dataset generalization, while acknowledging that broader generalization (e.g., across solar-cycle regimes or instrument domains) remains an important direction for future benchmarking.

Model performance was evaluated using standard regression metrics, including MAE, Root Mean Squared Error (RMSE), and the coefficient of determination ($R^2$), calculated separately for the training and test sets. These metrics allowed us to assess both the absolute and relative accuracy of the model in capturing daily sunspot variability from solar images alone. {Training was performed on a consumer-grade NVIDIA GPU (GeForce MX250, 2~GB VRAM) using TensorFlow/Keras, with typical training runs completing within a few hours depending on early stopping.}

\begin{figure*}[htb]
    \centering
    \includegraphics[width=0.65\textwidth]{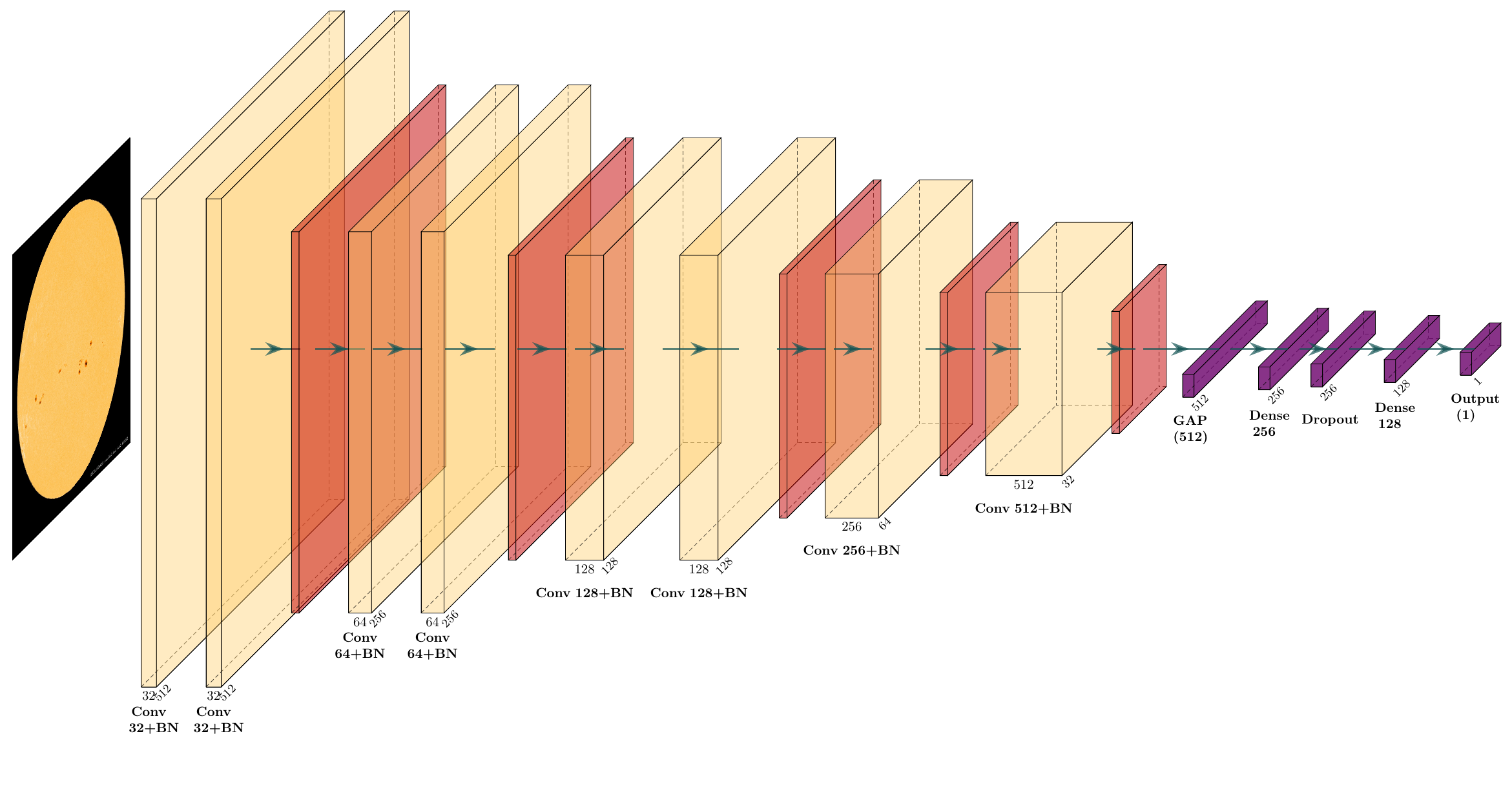}
    \caption{Schematic overview of the final CNN regression architecture. The model processes $512\times512$ three-channel HMI continuum quicklook images through stacked $3\times3$ convolutional layers with batch normalization (BN) and max pooling, followed by Global Average Pooling (GAP; 512) and a dense regression head (Dense 256 + dropout, Dense 128, Output 1).}
    \label{fig:cnn_arch}
\end{figure*}

{Figure~\ref{fig:cnn_arch} summarizes the final network topology. The backbone follows a VGG-like pattern of stacked $3\times3$ convolutions with batch normalization and periodic max-pooling, which progressively reduces the spatial resolution (from $512\times512$ down to $16\times16$) while increasing the number of feature channels (32 to 512). This design encourages the model to learn low-level contrast and edge cues in early layers and increasingly abstract, multi-scale representations of active-region morphology at deeper stages. A Global Average Pooling (GAP) layer then aggregates the final feature maps into a compact 512-dimensional descriptor, which is mapped to a scalar estimate through a lightweight fully connected head (Dense 256 with dropout, Dense 128, and a final linear neuron). This combination retains full-disk context, limits the number of parameters in the regressor head, and supports interpretability analyses based on gradient attribution.}

\section{Results and Evaluation}
\label{sec:results}

\subsection{Quantitative results, regression performance and residual analysis}

\begin{figure}[htb]
  \centering
\includegraphics[width=0.45\linewidth]{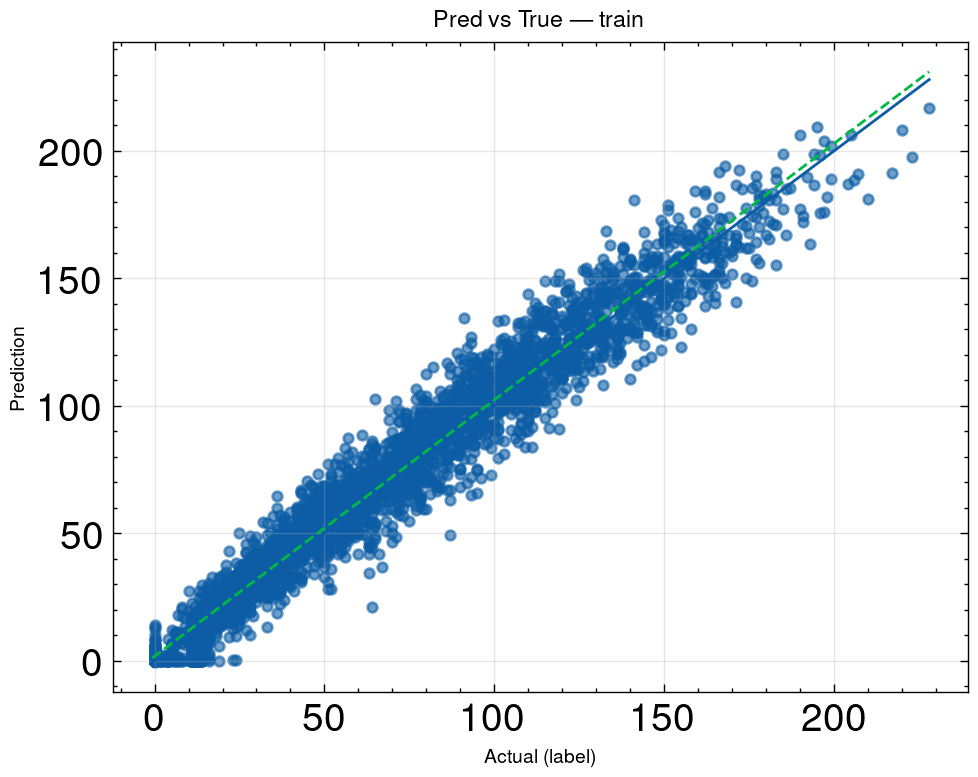}
\includegraphics[width=0.45\linewidth]{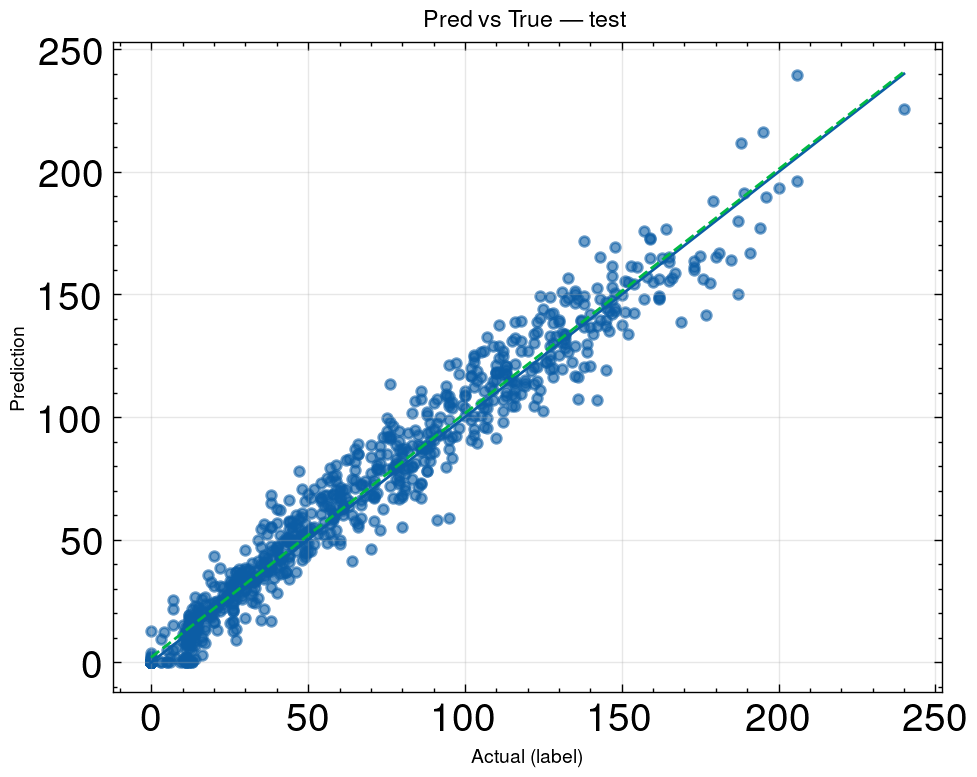}
\caption{Predicted versus observed daily sunspot numbers. Solid line: identity; dashed line: linear fit. The points cluster around the identity line for both splits, indicating strong agreement.}
\label{fig:pred_true}
\end{figure}

\begin{figure}[htb]
  \centering
\includegraphics[width=0.48\linewidth]{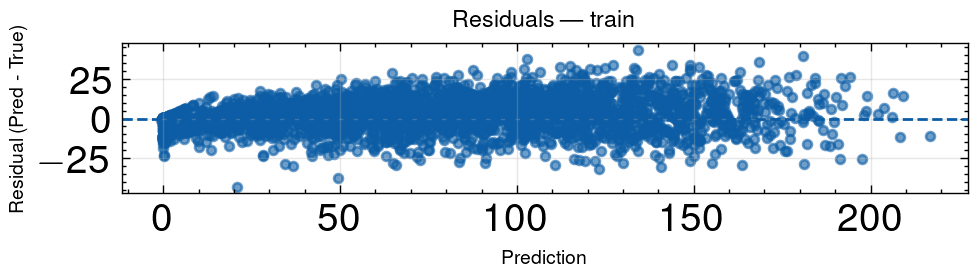}
\includegraphics[width=0.48\linewidth]{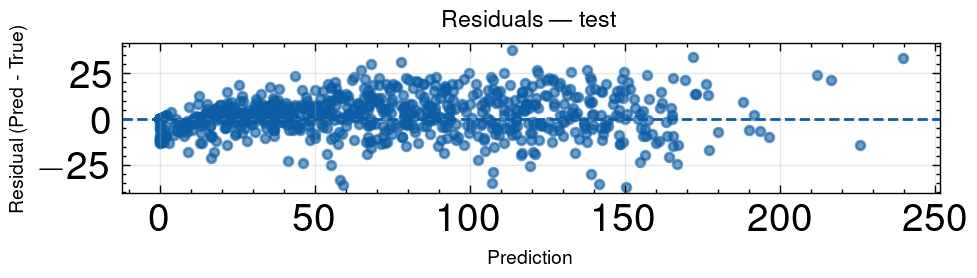}
\caption{Residuals (Predicted $-$ True) versus predicted value. Residuals are centered near zero with no strong heteroscedastic pattern; slight underestimation appears at the highest activity levels.}
  \label{fig:residuals}
\end{figure}

\begin{figure}[htb]
  \centering
\includegraphics[width=0.48\linewidth]{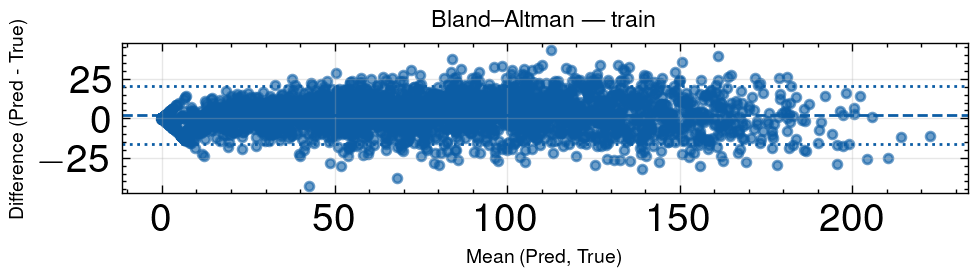}
\includegraphics[width=0.48\linewidth]{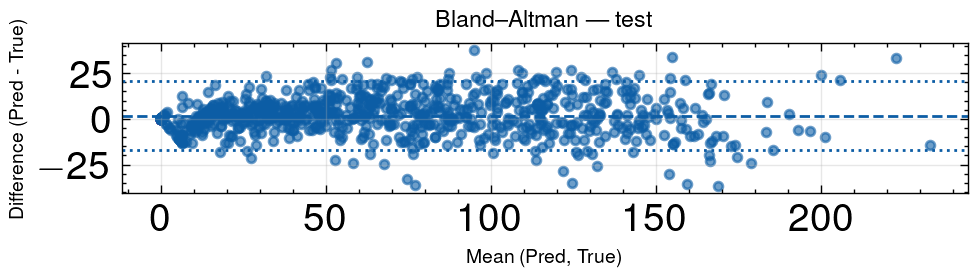}
\caption{Bland–Altman plots showing agreement between predictions and ground truth. Dashed line: mean difference (bias); dotted lines: limits of agreement ($\pm1.96\sigma$). Bias is close to zero; wider dispersion occurs for high sunspot activity.}
  \label{fig:bland_altman}
\end{figure}

\begin{figure}[htb]
  \centering
\includegraphics[width=0.48\linewidth]{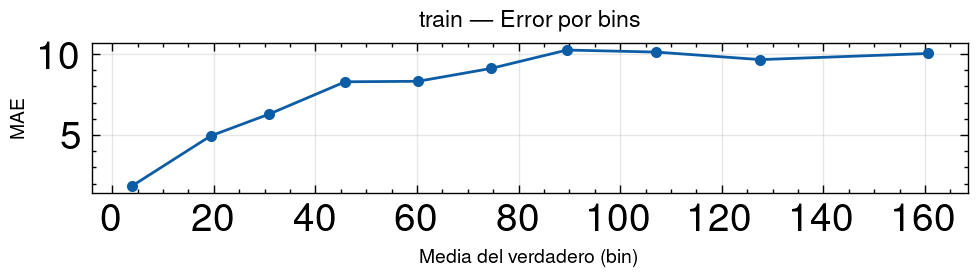}
\includegraphics[width=0.48\linewidth]{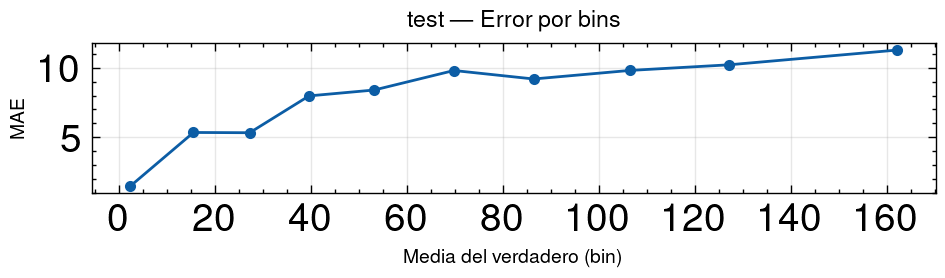}
\caption{Mean Absolute Error (MAE) computed within quantile bins of the true sunspot number. Errors increase with activity level, reflecting the broader dynamic range near solar maxima.}
  \label{fig:error_bins}
\end{figure}

The trained convolutional neural network demonstrated strong performance in estimating daily sunspot numbers from continuum solar images. Training curves showed rapid convergence within the first 15 epochs, followed by stabilization, with early stopping applied to prevent overfitting. Table~\ref{tab:metrics} summarizes the predictive performance of the proposed CNN on both training and test splits. We report standard regression metrics—MAE, RMSE, and $R^2$—together with Pearson’s $r$, Spearman’s $\rho$, and bias, each accompanied by 95\% bootstrap confidence intervals. The results indicate {high agreement with the SILSO ground truth, with an error structure that remains stable overall and increases moderately toward high-activity regimes, consistent with Figures~\ref{fig:pred_true}–\ref{fig:error_bins}.}

\begin{table}[htb]
\centering
\caption{Model performance with 95\% bootstrap confidence intervals.}
\label{tab:metrics}
\begin{tabular}{lcccccc}
\hline
\textbf{Split} & MAE & RMSE & $R^2$ & Pearson $r$ & Spearman $\rho$ & Bias \\
\hline
Train & 6.82 [6.61, 7.03] & 9.64 [9.38, 9.90] & 0.964 [0.961, 0.966] & 0.983 [0.982, 0.984] & 0.984 [0.982, 0.984] & 2.11 [1.84, 2.41] \\
Test  & 6.74 [6.29, 7.19] & 9.75 [9.18, 10.35] & 0.964 [0.960, 0.969] & 0.983 [0.981, 0.985] & 0.982 [0.979, 0.984] & 1.77 [1.15, 2.36] \\
\hline
\end{tabular}
\end{table}

{Both splits achieved comparable performance (Table~\ref{tab:metrics}), with overlapping confidence intervals across all metrics. On the test split, the model reached $R^2=0.964$ with Pearson’s $r=0.983$ and Spearman’s $\rho=0.982$, indicating strong agreement with the SILSO reference. The mean bias is slightly positive in both splits, suggesting a small overall tendency to overestimate sunspot numbers on average, while remaining within narrow confidence bounds.}

{Figure~\ref{fig:pred_true} shows predicted versus observed sunspot numbers for both training and test sets. The regression lines closely follow the identity line, indicating accurate calibration across most of the activity range. Although the model processes each daily image independently (i.e., it is not a temporal model), the time-ordered sequence of daily estimates reproduces the observed day-to-day variations of solar activity. Dispersion increases at higher sunspot numbers, consistent with the larger intrinsic variability during active periods.}

{Residual and Bland--Altman plots (Figures~\ref{fig:residuals} and \ref{fig:bland_altman}) further illustrate the model’s error structure. Residuals remain centered close to zero, but their spread increases with activity level, indicating mild heteroscedasticity. The Bland--Altman analysis shows a small positive mean difference (consistent with the overall bias), with wider limits of agreement at higher sunspot numbers. In the most active regimes, a slight regression-to-the-mean behavior is observed, with occasional underestimation of extreme values, likely reflecting the reduced availability of high-activity examples.}

{An error-stratified analysis by bins of solar activity (Figure~\ref{fig:error_bins}) shows that prediction error increases with the mean sunspot number. On the test split, MAE rises from $\approx 2$ at very low activity to $\approx 11$ in the highest-activity bin, consistent with both the broader dynamic range near solar maxima and the relative scarcity of extreme-activity days. Notably, the sign of the bias can change at the upper end, indicating a tendency to slightly underpredict the most extreme values while mildly overpredicting intermediate activity levels.}

{Taken together, these diagnostics indicate that the proposed CNN provides a well-calibrated mapping from full-disk continuum imagery to the daily sunspot number, with nearly identical behavior on training and test splits. The remaining discrepancies are largely structured rather than random: error variance increases with activity level and the most extreme cases exhibit a mild regression-to-the-mean effect, which is consistent with both the wider dynamical range near solar maxima and the smaller number of high-activity examples available for learning. Importantly, because the model operates on each daily image independently, the apparent time evolution in the outputs reflects the chronological ordering of independent estimates rather than an explicit temporal component in the architecture. These patterns motivate the interpretability analyses in the next section, where we examine which solar-disk regions and visual structures drive the model’s estimates across different activity regimes. Overall, the results support vision-based regression as a practical pathway to estimate classical heliophysical indices from full-disk imagery in a fully automated manner.}

\subsection{Model Interpretability}

To better understand the internal decision process of the convolutional neural network, we applied two complementary explainability techniques: Gradient-weighted Class Activation Mapping (Grad-CAM) and Integrated Gradients (IG) with SmoothGrad-IG. These methods provide visual insights into which regions of the input images contributed most strongly to the model’s predictions.

\begin{figure}[htb]
  \centering
  \includegraphics[width=0.24\linewidth]{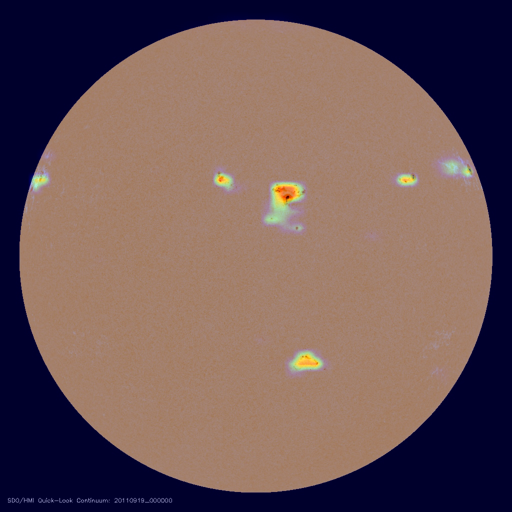}
  \includegraphics[width=0.24\linewidth]{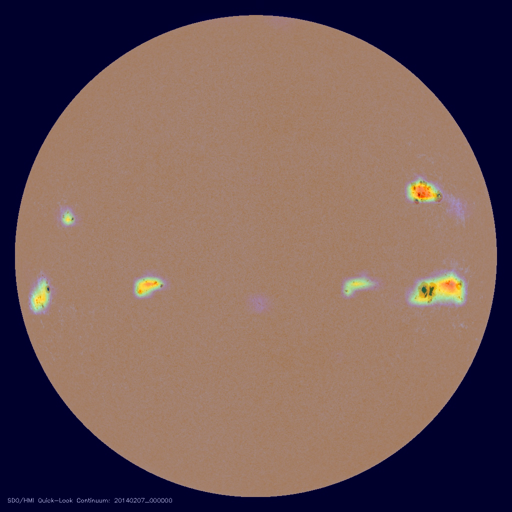}
  \includegraphics[width=0.24\linewidth]{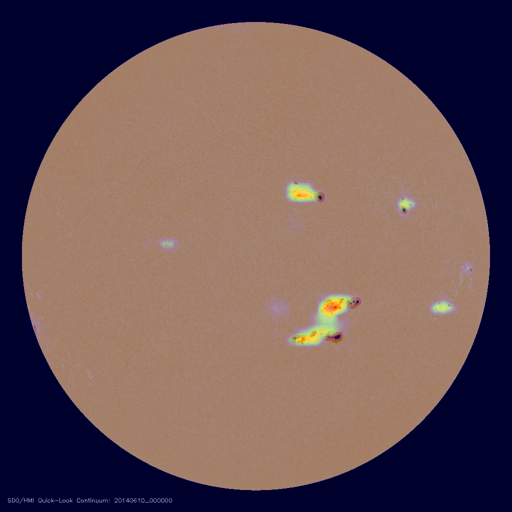}
  \includegraphics[width=0.24\linewidth]{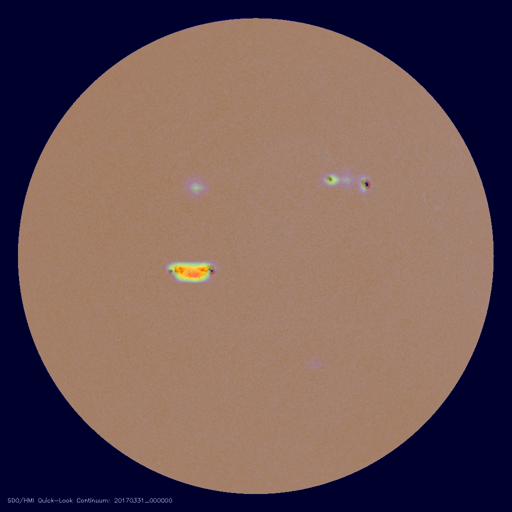} \\
  \includegraphics[width=0.24\linewidth]{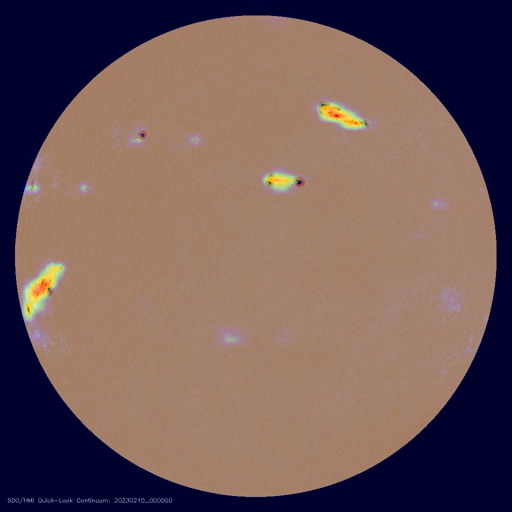}
  \includegraphics[width=0.24\linewidth]{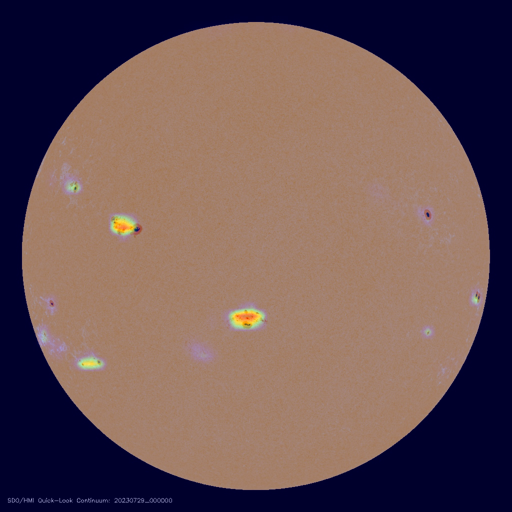}
  \includegraphics[width=0.24\linewidth]{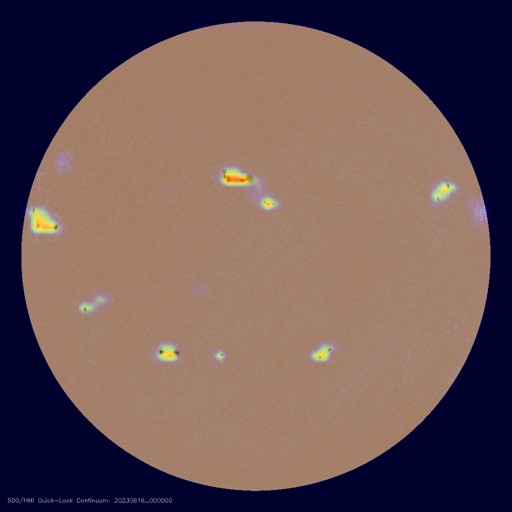}
  \includegraphics[width=0.24\linewidth]{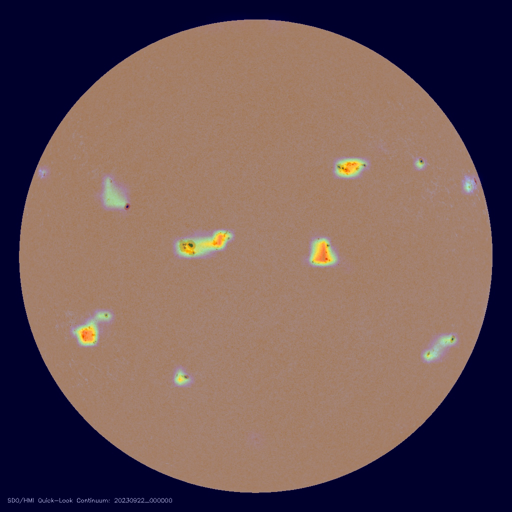}
  \caption{Grad-CAM overlays for representative HMI continuum quick-look images. Warmer colors indicate higher attribution to the predicted sunspot number. The examples correspond to (left-to-right, top-to-bottom): 2011--09--19, 2014--02--07, 2014--06--10, 2017--03--31, 2023--02--10, 2023--07--29, 2023--08--16, and 2023--09--22. In all cases, the strongest attribution concentrates on sunspot umbra/penumbra regions.}
  \label{fig:gradcam}
\end{figure}

{Figure~\ref{fig:gradcam} shows Grad-CAM overlays for representative dates spanning low to high solar activity. 
The highlighted regions concentrate on localized intensity depressions associated with sunspot umbrae and surrounding penumbrae, in agreement with the visual definition of sunspot-bearing structures in continuum images. 
Across all examples, the strongest attributions remain confined to compact active-region patches rather than the quiet photosphere or the image margins, supporting that the CNN relies primarily on physically meaningful morphology.}

\begin{figure}[htb]
  \centering
  \includegraphics[width=0.75\linewidth]{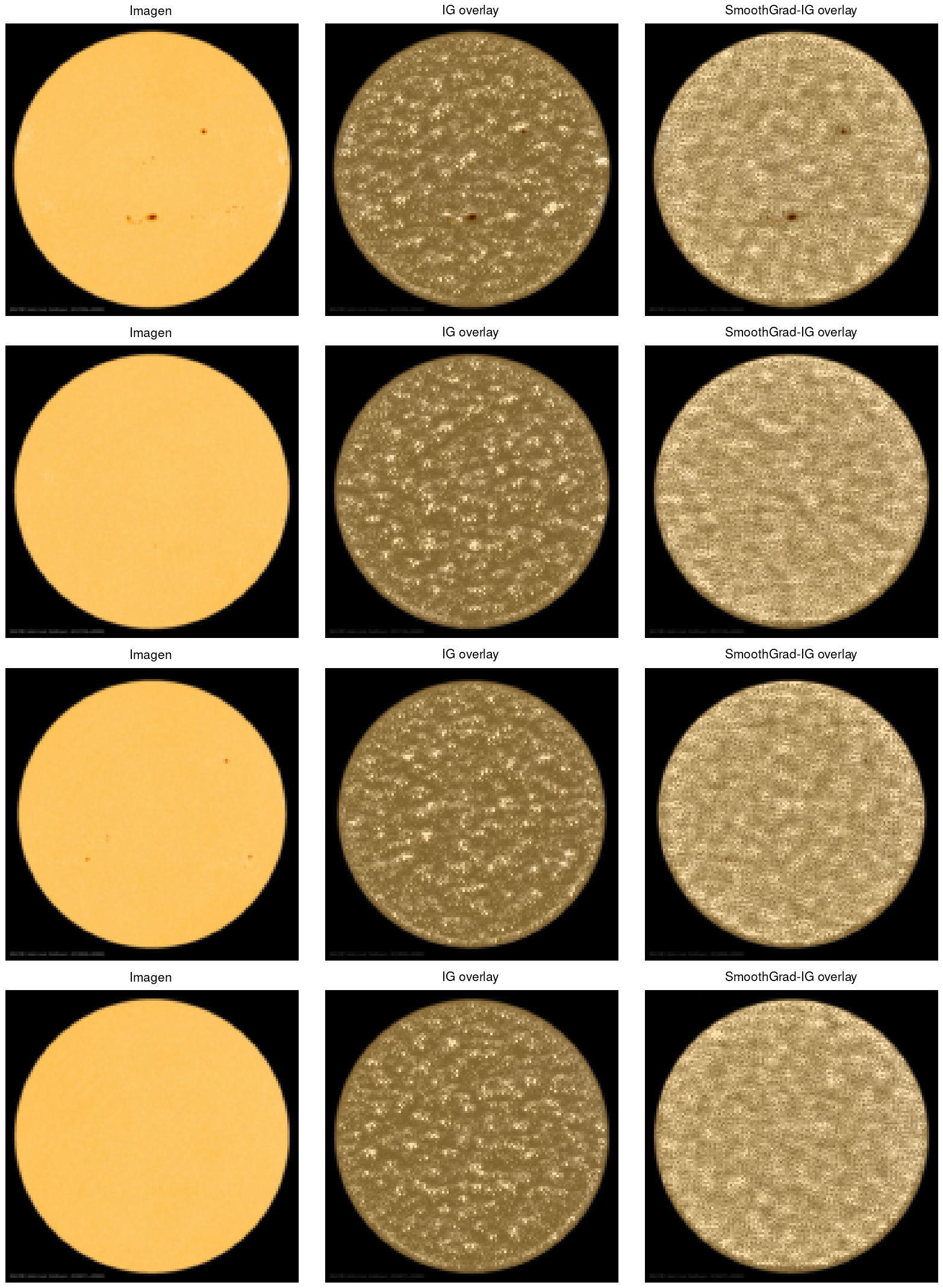}
  \caption{Integrated Gradients (IG) and SmoothGrad-IG attribution overlays for representative samples. Columns show the original image, IG overlay, and SmoothGrad-IG overlay. IG-based methods provide complementary, pixel-level relevance patterns; SmoothGrad-IG reduces high-frequency noise and reinforces the contribution of sunspot-bearing regions while also reflecting weaker sensitivity to global photometric context.}
  \label{fig:ig}
\end{figure}

{Complementary attribution maps from Integrated Gradients (IG) and SmoothGrad-IG are shown in Figure~\ref{fig:ig}. 
Compared with Grad-CAM (which is tied to convolutional feature maps and thus tends to localize salient regions), IG-based methods operate at the pixel level and often yield more spatially distributed responses on textured solar backgrounds. 
In our case, SmoothGrad-IG reduces high-frequency noise and consistently reinforces the relevance of sunspot-bearing areas, while also indicating a weaker sensitivity to broader photometric context (e.g., global contrast variations) that can modulate regression outputs. 
Accordingly, we interpret IG/SmoothGrad-IG as a complementary, qualitative check rather than a precise segmentation of sunspot pixels.}

{We emphasize that these explainability tools serve different purposes: Grad-CAM provides coarse but spatially localized evidence tied to deep convolutional representations, whereas IG variants can appear more diffuse on textured inputs because they attribute relevance at the pixel level. Using both perspectives helps validate that the dominant signal remains aligned with sunspot structures.}

{Overall, the explainability results support the physical plausibility of the learned mapping: the dominant attribution consistently aligns with sunspot structures in continuum intensity images, which are the direct visual drivers of the SILSO sunspot number.}

\section{Discussion and Conclusions}
\label{sec:conclusions}

{This work demonstrates that convolutional neural networks trained directly on solar continuum images can provide accurate estimates of daily sunspot numbers. By avoiding explicit feature engineering or manual counting, the proposed approach leverages the full spatial information of solar imagery and achieves competitive performance in an image-to-scalar regression setting. On the independent test split, our model reached $R^2 = 0.964$ with RMSE $= 9.75$ and MAE $= 6.74$, indicating strong agreement with the SILSO daily index across a wide activity range.}

{Table~\ref{tab:comparative} summarizes representative studies on sunspot-number modeling using statistical, machine learning, and deep learning approaches. The comparison should be interpreted with caution because many works address different targets (monthly means, smoothed indices, or multi-step estimation) and different input modalities (time series rather than images). Still, three broad trends emerge: (i) classical statistical forecasters (e.g., ARIMA/ETS variants) provide useful baselines but typically lag behind modern ML/DL methods in accuracy for long-horizon or multi-step settings; (ii) hybrid and ensemble pipelines can achieve high skill but often rely on decomposition steps and engineered representations; and (iii) deep learning models (convolutional and recurrent/attention-based) consistently report strong performance, especially on aggregated or smoothed sunspot indices. Within this landscape, our contribution is complementary: a direct, vision-based mapping from full-disk imagery to the daily sunspot number.}

Our CNN provides a competitive baseline for direct image-based estimation while keeping the pipeline simple and fully end-to-end. Relative to image-based feature-engineering approaches (e.g., topological descriptors followed by classical regressors), the present model removes the need to define handcrafted image features and enables interpretability analyses directly on the learned representations. At the same time, we emphasize that many recent high-$R^2$ reports in the literature correspond to monthly or smoothed time-series forecasting tasks, which are not directly comparable to daily image-to-scalar estimation.

\begin{table}[htb]
\centering
\caption{Comparative summary of sunspot number prediction studies. $R^2$ and RMSE are reported when available. The last row contains our results for direct image-based prediction (SDO/HMI + SILSO v2.0).}
\label{tab:comparative}
\begin{tabular}{p{0.8cm} p{3.2cm} p{3.2cm} p{3cm} p{3cm} p{2cm}}
\hline
\textbf{Year} & \textbf{Method / Model} & \textbf{Dataset} & \textbf{$R^2$} & \textbf{RMSE} & \textbf{Reference} \\
\hline
2022 & SARIMA, ES, Prophet, LSTM/GRU, Transformer, Informer; XGBoost-DL (best) & SSN monthly (historical) & -- & RMSE = 25.70 and MAE = 19.82 & \citep{2022arXiv220305757D} \\
2024 & GRU + SMA + STL, pinball loss (probabilistic) & SILSO (sunspot series) & one-step: 0.9373, two-step: 0.9351, three-step: 0.9355 & one-step: 19.0557, two-step: 19.3829, three-step: 19.3117 &  \cite{cui2024probabilistic}\\
2024 & 1D-CNN + BiLSTM + Multi-Head Attention & Monthly SSN (1811–2022) & Corr (1-month) = 0.9996 ($R^2$ $\approx$ 0.9992), Corr (5-months) = 0.9981 ($R^2$ $\approx$ 0.9992), Corr (7-month) = 0.9981 ($R^2$ $\approx$ 0.9956) & RMSE (1-month) = 2.07, RMSE (5-months) = 2.89, RMSE (7-months) = 3.39 & \cite{chen2024prediction} \\
2024 & Combinatorial deep-learning (hybrid blocks) & SSN (multiple cycles) & -- & 17.13 & \cite{su2024solar} \\
2024 & SARIMA + RF (hybrid), STL comparisons & Monthly SSN (1749–2023) & -- & Best 0.85 (Scaled data) &  \cite{xu2024data}\\
2024 & Spectral analysis + ML (RF, SVM, etc.) & Hemispheric SSN (SILSO) & -- & Northern hemisphere (6.1) and the Southern hemisphere (6.8) & \cite{rodriguez2024hemispheric} \\
2024 & Topological feature extraction + ML regression & SDO/HMI images + SILSO & Best model ExtraTrees $R^2$ = 0.97 & Best model ExtraTrees RMSE = 9.14 & \cite{20244800857S} \\
2024 & Hybrid CNN + LSTM/GRU / ensemble & Monthly SSN / 13-month smoothed & 0.9585 & 1.64 & \cite{2024AdSpR734342K} \\
2024 & CEEMDAN (decomposition) + GRU + error correction & Monthly SSN (recent) & Corr (1-setp) = 0.99971 → $R^2$ $\approx$ 0.99942 & RMSE = 1.319 & \cite{yang2024hybrid} \\
2025 & SARIMA, GPR, LightGBM, LSTM (comparative) & Monthly mean SSN (long record) & Best model: SARIMA 1 step ahead. $R^2$ = 0.94 & Best model: SARIMA 1 step ahead. RMSE = 14.37 & \citep{paraskakis2025prediction} \\
2025 & LSTM + WGAN (data augmentation) & SSN (monthly, long record) & 0.977 & 5.02 & \citep{2025FrASS1241299Y} \\
\textbf{2025} & \textbf{CNN (vision-based, direct images)} & \textbf{SDO/HMI + SILSO v2.0 (daily)} & \textbf{0.964} & \textbf{9.75} & This work \\
\hline
\end{tabular}
\end{table}
\noindent

{Future work should focus on three directions. First, a synergistic multimodal model would combine an image encoder (e.g., a CNN/ViT operating on HMI continuum images) with complementary temporal predictors (e.g., recent sunspot-number history and other scalar proxies such as F10.7), using feature fusion (early or late fusion) to capture both instantaneous morphology and short-term temporal context. Second, systematic benchmarking across cycles should be pursued via cross-cycle holdouts (e.g., training on Solar Cycle 24 and testing on Cycle 25, and vice versa) and rolling-origin evaluations, to quantify robustness under domain shifts and changing activity regimes. Extending benchmarks across multiple solar cycles requires not only long sunspot-number records but also consistent full-disk imagery. While SDO/HMI provides homogeneous observations from 2010 onward, earlier cycles would require incorporating pre-SDO image archives, such as SoHO-era observations (e.g., MDI continuum/intensity products) or long-term ground-based full-disk programs. A key challenge is cross-instrument domain shift: differences in spatial resolution, photometric response, compression and rendering pipelines, cadence, and data gaps can strongly affect image statistics. Addressing this would likely require careful harmonization (disk normalization, consistent cropping/rescaling, intensity standardization) and explicit cross-instrument evaluation (train on one instrument, test on another), potentially supported by domain adaptation methods. Third, interpretability tools (Grad-CAM, IG) can be expanded into routine quality-control diagnostics, supporting trust and scientific plausibility checks in automated monitoring pipelines.}

{In summary, our results show that direct, vision-based regression from full-disk continuum images to the daily SILSO sunspot number is feasible and accurate, providing a practical baseline for integrating modern computer vision into automated sunspot-number estimation.}

\section*{Acknowledgments}

The authors acknowledge the Solar Influences Data Analysis Center (SILSO), Royal Observatory of Belgium, for providing the sunspot number series, and NASA’s Solar Dynamics Observatory (SDO)/HMI team for making solar images publicly available. We are also grateful to the NMDB database (www.nmdb.eu), funded under the European Union’s FP7 programme (contract no. 213007), for supporting heliophysical data services. 

This work was supported by the Research Directorate of Universidad Tecnológica de Bolívar (UTB), Cartagena, Colombia, which provided institutional support and encouragement during the development of this study.

\bibliography{main}

@ARTICLE{2022SoPh..297..158H,
       author = {{Hanaoka}, Yoichiro},
        title = "{Automated Sunspot Detection as an Alternative to Visual Observations}",
      journal = {\solphys},
         year = 2022,
        month = dec,
       volume = {297},
       number = {12},
          eid = {158},
        pages = {158},
          doi = {10.1007/s11207-022-02089-z},
archivePrefix = {arXiv},
       eprint = {2211.13552},
 primaryClass = {astro-ph.SR},
       adsurl = {https://ui.adsabs.harvard.edu/abs/2022SoPh..297..158H}
}

@ARTICLE{2024AJ....167...52Z,
       author = {{Zhao}, Cui and {Yang}, Shangbin and {Wang}, Tingmei and {Zhao}, Haiyan and {Liu}, Shiyuan and {He}, Fangyuan and {Hu}, Zhengkun},
        title = "{An Automatic Approach for Grouping Sunspots and Calculating Relative Sunspot Number on SDO/HMI Continuum Images}",
      journal = {\aj},
         year = 2024,
        month = feb,
       volume = {167},
       number = {2},
          eid = {52},
        pages = {52},
          doi = {10.3847/1538-3881/ad11e2},
archivePrefix = {arXiv},
       eprint = {2401.08949},
 primaryClass = {astro-ph.SR},
       adsurl = {https://ui.adsabs.harvard.edu/abs/2024AJ....167...52Z}
}

@ARTICLE{2012EPJP..127...43C,
       author = {{Chattopadhyay}, Goutami and {Chattopadhyay}, Surajit},
        title = "{Monthly sunspot number time series analysis and its modeling through autoregressive artificial neural network}",
      journal = {European Physical Journal Plus},
         year = 2012,
        month = apr,
       volume = {127},
       number = {4},
          eid = {43},
        pages = {43},
          doi = {10.1140/epjp/i2012-12043-9},
archivePrefix = {arXiv},
       eprint = {1204.3991},
 primaryClass = {physics.gen-ph},
       adsurl = {https://ui.adsabs.harvard.edu/abs/2012EPJP..127...43C}
}

@ARTICLE{2002A&A...386..313O,
       author = {{Orfila}, A. and {Ballester}, J.~L. and {Oliver}, R. and {Alvarez}, A. and {Tintor{\'e}}, J.},
        title = "{Forecasting the solar cycle with genetic algorithms}",
      journal = {\aap},
         year = 2002,
        month = apr,
       volume = {386},
        pages = {313-318},
          doi = {10.1051/0004-6361:20020246},
       adsurl = {https://ui.adsabs.harvard.edu/abs/2002A&A...386..313O}
}

@ARTICLE{2025RASTI...4...24M,
       author = {{Meadows}, P J},
        title = "{SOHO/MDI and SDO/HMI sunspot area measurement and analysis}",
      journal = {RAS Techniques and Instruments},
         year = 2025,
        month = jan,
       volume = {4},
          eid = {rzaf024},
        pages = {rzaf024},
          doi = {10.1093/rasti/rzaf024},
       adsurl = {https://ui.adsabs.harvard.edu/abs/2025RASTI...4...24M}
}

@ARTICLE{2024SoPh..299..156H,
       author = {{Hanaoka}, Yoichiro},
        title = "{Evaluation of Sunspot Areas Derived by Automated Sunspot-Detection Methods}",
      journal = {\solphys},
         year = 2024,
        month = nov,
       volume = {299},
       number = {11},
          eid = {156},
        pages = {156},
          doi = {10.1007/s11207-024-02402-y},
archivePrefix = {arXiv},
       eprint = {2411.00415},
 primaryClass = {astro-ph.SR},
       adsurl = {https://ui.adsabs.harvard.edu/abs/2024SoPh..299..156H}
}

@article{cedazo2020improving,
  title={Improving the results of citizen science projects through reputation systems: The case of Wolf’s Number Experiment},
  author={Cedazo, Raquel and Gonzalez, Esteban and Serra-Ricart, Miquel and Brunete, Alberto},
  journal={IEEE Access},
  volume={8},
  pages={186026--186038},
  year={2020},
  publisher={IEEE},
doi={10.1109/ACCESS.2020.3030006}
}

@article{sarsembayeva2021detecting,
  title={Detecting the Sun’s active region using image processing techniques},
  author={Sarsembayeva, A and Odsuren, M and Belisarova, F and Sarsembay, A and Maftunzada, SAL},
  journal={Physical Sciences and Technology},
  volume={8},
  number={3-4},
  pages={48--53},
  year={2021},
doi={10.26577/phst.2021.v8.i2.07}
}

@article{sarsembayeva2025solar,
  title={Solar Magnetic Activity and Its Terrestrial Impact through Correlations with Drought Indices},
  author={Sarsembayeva, Aiganym and Ryssaliyeva, Laura and Belissarova, Farida and Sarsembay, Akmaral},
  journal={Physical Sciences and Technology},
  volume={12},
  number={1-2},
  pages={38--44},
  year={2025},
doi={10.26577/phst20251214}
}

@ARTICLE{1990ApJ...356..733B,
       author = {{Bornmann}, P.~L.},
        title = "{Limits to Derived Flare Properties Using Estimates for the Background Fluxes: Examples from GOES}",
      journal = {\apj},
         year = 1990,
        month = jun,
       volume = {356},
        pages = {733},
          doi = {10.1086/168880},
       adsurl = {https://ui.adsabs.harvard.edu/abs/1990ApJ...356..733B}
}

@ARTICLE{2001MNRAS.323..223B,
       author = {{Baranyi}, T. and {Gyori}, L. and {Ludm{\'a}ny}, A. and {Coffey}, H.~E.},
        title = "{Comparison of sunspot area data bases}",
      journal = {\mnras},
         year = 2001,
        month = may,
       volume = {323},
       number = {1},
        pages = {223-230},
          doi = {10.1046/j.1365-8711.2001.04195.x},
       adsurl = {https://ui.adsabs.harvard.edu/abs/2001MNRAS.323..223B}
}

@ARTICLE{2024arXiv240502545S,
       author = {{Sakpal}, Shlesh},
        title = "{Prediction of Space Weather Events through Analysis of Active Region Magnetograms using Convolutional Neural Network}",
      journal = {arXiv e-prints},
         year = 2024,
        month = may,
          eid = {arXiv:2405.02545},
        pages = {arXiv:2405.02545},
          doi = {10.48550/arXiv.2405.02545},
archivePrefix = {arXiv},
       eprint = {2405.02545},
 primaryClass = {astro-ph.SR},
       adsurl = {https://ui.adsabs.harvard.edu/abs/2024arXiv240502545S}
}

@inproceedings{rong2014improved,
  title={An improved CANNY edge detection algorithm},
  author={Rong, Weibin and Li, Zhanjing and Zhang, Wei and Sun, Lining},
  booktitle={2014 IEEE international conference on mechatronics and automation},
  pages={577--582},
  year={2014},
  organization={IEEE},
doi={10.1109/ICMA.2014.6885761}
}

@incollection{cheng2012improved,
  title={An improved Canny edge detection algorithm},
  author={Cheng, You-e},
  booktitle={Recent Advances in Computer Science and Information Engineering: Volume 3},
  pages={551--558},
  year={2012},
  publisher={Springer},
doi={10.1007/978-3-642-25766-7_73}
}

@article{love2020analyzing,
  title={Analyzing AIA flare observations using convolutional neural networks},
  author={Love, Teri and Neukirch, Thomas and Parnell, Clare E},
  journal={Frontiers in Astronomy and Space Sciences},
  volume={7},
  pages={34},
  year={2020},
  publisher={Frontiers Media SA},
doi={10.3389/fspas.2020.00034}
}

@article{chola2022detection,
  title={Detection and classification of sunspots via deep convolutional neural network},
  author={Chola, Channabasava and Benifa, JV Biabl},
  journal={Global Transitions Proceedings},
  volume={3},
  number={1},
  pages={177--182},
  year={2022},
  publisher={Elsevier},
doi={10.1016/j.gltp.2022.03.006}
}

@article{diaz2022towards,
  title={Towards the identification and classification of solar granulation structures using semantic segmentation},
  author={D{\'\i}az Castillo, SM and Asensio Ramos, A and Fischer, CE and Berdyugina, SV},
  journal={Frontiers in Astronomy and Space Sciences},
  volume={9},
  pages={896632},
  year={2022},
  publisher={Frontiers Media SA},
doi={10.3389/fspas.2022.896632}
}

@ARTICLE{2020FrP.....8...45F,
       author = {{Feng}, Li and {Gan}, Weiqun and {Liu}, Siqing and {Wang}, Huaning and {Li}, Hui and {Xu}, Long and {Zong}, Weiguo and {Zhang}, Xiaoxing and {Zhu}, Yaguang and {Wu}, Haiyan and {Chen}, Anqin and {Cui}, Yanmei and {Dai}, Xinghua and {Guo}, Juan and {He}, Han and {Huang}, Xin and {Lu}, Lei and {Song}, Qiao and {Wang}, Jingjing and {Zhong}, Qiuzhen and {Chen}, Ling and {Du}, Zhanle and {Guo}, Xingliang and {Huang}, Yu and {Li}, Hu and {Li}, Ying and {Xiong}, Senlin and {Yang}, Shenggao and {Ying}, Beili},
        title = "{Space Weather Related to Solar Eruptions with the ASO-S Mission}",
      journal = {Frontiers in Physics},
         year = 2020,
        month = mar,
       volume = {8},
          eid = {45},
        pages = {45},
          doi = {10.3389/fphy.2020.00045},
       adsurl = {https://ui.adsabs.harvard.edu/abs/2020FrP.....8...45F}
}

@ARTICLE{2019AdAst2019E..27F,
       author = {{Fang}, Yuanhui and {Cui}, Yanmei and {Ao}, Xianzhi},
        title = "{Deep Learning for Automatic Recognition of Magnetic Type in Sunspot Groups}",
      journal = {Advances in Astronomy},
         year = 2019,
        month = aug,
       volume = {2019},
          eid = {9196234},
        pages = {9196234},
          doi = {10.1155/2019/9196234},
       adsurl = {https://ui.adsabs.harvard.edu/abs/2019AdAst2019E..27F}
}

@INPROCEEDINGS{2023aike.conf...83P,
       author = {{Pandey}, Chetraj and {Ji}, Anli and {Angryk}, Rafal A. and {Aydin}, Berkay},
        title = "{Towards Interpretable Solar Flare Prediction with Attention-based Deep Neural Networks}",
    booktitle = {2023 IEEE Sixth International Conference on Artificial Intelligence and Knowledge Engineering},
         year = 2023,
        month = sep,
        pages = {83-90},
          doi = {10.1109/AIKE59827.2023.00021},
archivePrefix = {arXiv},
       eprint = {2309.04558},
 primaryClass = {cs.LG},
       adsurl = {https://ui.adsabs.harvard.edu/abs/2023aike.conf...83P}
}

@ARTICLE{2020ApJ...891...10L,
       author = {{Li}, Xuebao and {Zheng}, Yanfang and {Wang}, Xinshuo and {Wang}, Lulu},
        title = "{Predicting Solar Flares Using a Novel Deep Convolutional Neural Network}",
      journal = {\apj},
         year = 2020,
        month = mar,
       volume = {891},
       number = {1},
          eid = {10},
        pages = {10},
          doi = {10.3847/1538-4357/ab6d04},
       adsurl = {https://ui.adsabs.harvard.edu/abs/2020ApJ...891...10L}
}

@inproceedings{shen2024deep,
  title={Deep Computer Vision for Solar Physics Big Data: Opportunities and Challenges [Vision Paper]},
  author={Shen, Bo and Marena, Marco and Li, Chenyang and Li, Qin and Jiang, Haodi and Du, Mengnan and Xu, Jiajun and Wang, Haimin},
  booktitle={2024 IEEE International Conference on Big Data (BigData)},
  pages={1860--1864},
  year={2024},
  organization={IEEE},
doi={10.1109/BigData62323.2024.10825648},
archivePrefix = {arXiv},
       eprint = {2409.04850}
}

@article{xu2025solar,
  title={Solar flare forecasting using hybrid neural networks},
  author={Xu, Dan and Sun, Pengchao and Feng, Song and Liang, Bo and Dai, Wei},
  journal={The Astrophysical Journal Supplement Series},
  volume={276},
  number={2},
  pages={68},
  year={2025},
  publisher={IOP Publishing},
doi={10.3847/1538-4365/ada281}
}

@ARTICLE{2018ApJ...856....7H,
       author = {{Huang}, Xin and {Wang}, Huaning and {Xu}, Long and {Liu}, Jinfu and {Li}, Rong and {Dai}, Xinghua},
        title = "{Deep Learning Based Solar Flare Forecasting Model. I. Results for Line-of-sight Magnetograms}",
      journal = {\apj},
         year = 2018,
        month = mar,
       volume = {856},
       number = {1},
          eid = {7},
        pages = {7},
          doi = {10.3847/1538-4357/aaae00},
       adsurl = {https://ui.adsabs.harvard.edu/abs/2018ApJ...856....7H}
}

@ARTICLE{2022Univ....8..656L,
       author = {{Li}, Siqi and {Yuan}, Guowu and {Chen}, Jian and {Tan}, Chengming and {Zhou}, Hao},
        title = "{Self-Supervised Learning for Solar Radio Spectrum Classification}",
      journal = {Universe},
         year = 2022,
        month = dec,
       volume = {8},
       number = {12},
          eid = {656},
        pages = {656},
          doi = {10.3390/universe8120656},
archivePrefix = {arXiv},
       eprint = {2502.03778},
 primaryClass = {astro-ph.IM},
       adsurl = {https://ui.adsabs.harvard.edu/abs/2022Univ....8..656L}
}

@ARTICLE{2018A&A...618A.165Z,
       author = {{Zhang}, P.~J. and {Wang}, C.~B. and {Ye}, L.},
        title = "{A type III radio burst automatic analysis system and statistic results for a half solar cycle with Nan{\c{c}}ay Decameter Array data}",
      journal = {\aap},
         year = 2018,
        month = oct,
       volume = {618},
          eid = {A165},
        pages = {A165},
          doi = {10.1051/0004-6361/201833260},
archivePrefix = {arXiv},
       eprint = {1810.02921},
 primaryClass = {astro-ph.SR},
       adsurl = {https://ui.adsabs.harvard.edu/abs/2018A&A...618A.165Z}
}

@inproceedings{ji2023interpretable,
  title={Interpretable solar flare prediction with sliding window multivariate time series forests},
  author={Ji, Anli and Aydin, Berkay},
  booktitle={2023 IEEE International Conference on Big Data (BigData)},
  pages={1519--1524},
  year={2023},
  organization={IEEE},
doi={10.1109/BigData59044.2023.10386908}
}

@ARTICLE{2025arXiv250616194B,
       author = {{Bauer}, Maike and {Le Lou{\"e}dec}, Justin and {Amerstorfer}, Tanja and {Barnard}, Luke and {Barnes}, David and {Lammer}, Helmut},
        title = "{Solar Transient Recognition Using Deep Learning (STRUDL) for heliospheric imager data}",
      journal = {arXiv e-prints},
         year = 2025,
        month = jun,
          eid = {arXiv:2506.16194},
        pages = {arXiv:2506.16194},
          doi = {10.48550/arXiv.2506.16194},
archivePrefix = {arXiv},
       eprint = {2506.16194},
 primaryClass = {astro-ph.SR},
       adsurl = {https://ui.adsabs.harvard.edu/abs/2025arXiv250616194B}
}

@inproceedings{jha2019resunet++,
  title={Resunet++: An advanced architecture for medical image segmentation},
  author={Jha, Debesh and Smedsrud, Pia H and Riegler, Michael A and Johansen, Dag and De Lange, Thomas and Halvorsen, P{\aa}l and Johansen, H{\aa}vard D},
  booktitle={2019 IEEE international symposium on multimedia (ISM)},
  pages={225--2255},
  year={2019},
  organization={IEEE},
doi={10.1109/ISM46123.2019.00049}
}

@inproceedings{ronneberger2015u,
  title={U-net: Convolutional networks for biomedical image segmentation},
  author={Ronneberger, Olaf and Fischer, Philipp and Brox, Thomas},
  booktitle={International Conference on Medical image computing and computer-assisted intervention},
  pages={234--241},
  year={2015},
  organization={Springer},
doi={10.1007/978-3-319-24574-4_28}
}

@article{sierra2024predicting,
  title={Predicting sunspot number from topological features in spectral images I: Machine learning approach},
  author={Acevedo, DD Herrera},
  journal={Astronomy and Computing},
  volume={48},
  pages={100857},
  year={2024},
  publisher={Elsevier},
doi={10.1016/j.ascom.2024.100857}
}

@inproceedings{mercioni2020p,
  title={P-swish: Activation function with learnable parameters based on swish activation function in deep learning},
  author={Mercioni, Marina Adriana and Holban, Stefan},
  booktitle={2020 International Symposium on Electronics and Telecommunications (ISETC)},
  pages={1--4},
  year={2020},
  organization={IEEE},
doi={10.1109/ISETC50328.2020.9301059}
}

@ARTICLE{2024arXiv240708232R,
       author = {{Rahman}, Jamshaid Ul and {Zulfiqar}, Rubiqa and {Khan}, Asad and {Nimra}},
        title = "{SwishReLU: A Unified Approach to Activation Functions for Enhanced Deep Neural Networks Performance}",
      journal = {arXiv e-prints},
         year = 2024,
        month = jul,
          eid = {arXiv:2407.08232},
        pages = {arXiv:2407.08232},
          doi = {10.48550/arXiv.2407.08232},
archivePrefix = {arXiv},
       eprint = {2407.08232},
 primaryClass = {cs.LG},
       adsurl = {https://ui.adsabs.harvard.edu/abs/2024arXiv240708232R}
}

@article{gupta2020robust,
  title={Robust regularized extreme learning machine with asymmetric Huber loss function},
  author={Gupta, Deepak and Hazarika, Barenya Bikash and Berlin, Mohanadhas},
  journal={Neural Computing and Applications},
  volume={32},
  number={16},
  pages={12971--12998},
  year={2020},
  publisher={Springer},
doi={10.1007/s00521-020-04741-w}
}

@ARTICLE{2019arXiv190409237R,
       author = {{Reddi}, Sashank J. and {Kale}, Satyen and {Kumar}, Sanjiv},
        title = "{On the Convergence of Adam and Beyond}",
      journal = {arXiv e-prints},
         year = 2019,
        month = apr,
          eid = {arXiv:1904.09237},
        pages = {arXiv:1904.09237},
          doi = {10.48550/arXiv.1904.09237},
archivePrefix = {arXiv},
       eprint = {1904.09237},
 primaryClass = {cs.LG},
       adsurl = {https://ui.adsabs.harvard.edu/abs/2019arXiv190409237R}
}

@article{cui2024probabilistic,
  title={Probabilistic sunspot predictions with a gated recurrent units-based combined model guided by pinball loss},
  author={Cui, Zhesen and Ding, Zhe and Xu, Jing and Zhang, Shaotong and Wu, Jinran and Lian, Wei},
  journal={Scientific Reports},
  volume={14},
  number={1},
  pages={13601},
  year={2024},
  publisher={Nature Publishing Group UK London},
doi={10.1038/s41598-024-63878-z}
}

@article{chen2024prediction,
  title={Prediction of Sunspot Number with Hybrid Model Based on 1D-CNN, BiLSTM and Multi-Head Attention Mechanism.},
  author={Chen, Huirong and Liu, Song and Yang, Ximing and Zhang, Xinggang and Yang, Jianzhong and Fan, Shaofen},
  journal={Electronics (2079-9292)},
  volume={13},
  number={14},
  year={2024},
doi={10.3390/electronics13142804}
}

@article{su2024solar,
  title={Solar cycle prediction using a combinatorial deep learning model},
  author={Su, Xu and Liang, Bo and Feng, Song and Cai, Yunfang and Dai, Wei and Yang, Yunfei},
  journal={Monthly Notices of the Royal Astronomical Society},
  volume={527},
  number={3},
  pages={5675--5682},
  year={2024},
  publisher={Oxford University Press},
doi={10.1093/mnras/stad3451}
}

@article{xu2024data,
  title={Data-driven forecasting of sunspot cycles: pros and cons of a hybrid approach},
  author={Xu, Qinglin and Jain, Rekha and Xing, Wei},
  journal={Solar Physics},
  volume={299},
  number={2},
  pages={25},
  year={2024},
  publisher={Springer},
doi={10.1007/s11207-024-02270-6}
}

@article{rodriguez2024hemispheric,
  title={Hemispheric sunspot number prediction for solar cycles 25 and 26 using spectral analysis and machine learning techniques},
  author={Rodr{\'\i}guez, Jos{\'e}-V{\'\i}ctor and S{\'a}nchez Carrasco, V{\'\i}ctor Manuel and Rodr{\'\i}guez-Rodr{\'\i}guez, Ignacio and P{\'e}rez Aparicio, Alejandro Jes{\'u}s and Vaquero, Jos{\'e} Manuel},
  journal={Solar Physics},
  volume={299},
  number={8},
  pages={116},
  year={2024},
  publisher={Springer},
doi={10.1007/s11207-024-02363-2}
}

@ARTICLE{20244800857S,
       author = {{Sierra-Porta}, D. and {Tarazona-Alvarado}, M. and {Acevedo}, D.~D. Herrera},
        title = "{Predicting sunspot number from topological features in spectral images I: Machine learning approach}",
      journal = {Astronomy and Computing},
         year = 2024,
        month = jul,
       volume = {48},
          eid = {100857},
        pages = {100857},
          doi = {10.1016/j.ascom.2024.100857},
       adsurl = {https://ui.adsabs.harvard.edu/abs/2024A&C....4800857S}
}

@ARTICLE{2024AdSpR734342K,
       author = {{Kumar}, Abhijeet and {Kumar}, Vipin},
        title = "{Forecast of solar cycle 25 based on Hybrid CNN-Bidirectional-GRU (CNN-BiGRU) model and Novel Gradient Residual Correction (GRC) technique}",
      journal = {Advances in Space Research},
         year = 2024,
        month = apr,
       volume = {73},
       number = {8},
        pages = {4342-4362},
          doi = {10.1016/j.asr.2024.01.019},
       adsurl = {https://ui.adsabs.harvard.edu/abs/2024AdSpR..73.4342K}
}

@article{yang2024hybrid,
  title={A hybrid model based on CEEMDAN-GRU and error compensation for predicting sunspot numbers},
  author={Yang, Jianzhong and Liu, Song and Xuan, Shili and Chen, Huirong},
  journal={Electronics},
  volume={13},
  number={10},
  pages={1904},
  year={2024},
  publisher={MDPI},
doi={10.3390/electronics13101904}
}

@article{paraskakis2025prediction,
  title={Prediction of yearly mean sunspot number using machine learning methods},
  author={Paraskakis, Nikolaos and Hristopulos, Dionissios T},
  journal={Stochastic Environmental Research and Risk Assessment},
  pages={1--28},
  year={2025},
  publisher={Springer},
doi={10.1007/s00477-025-03024-x}
}

@ARTICLE{2025FrASS1241299Y,
       author = {{Yang}, Hao and {Zuo}, Pingbing and {Zhang}, Kun and {Shen}, Zhenning and {Zou}, Zhengyang and {Feng}, Xueshang},
        title = "{Forecasting long-term sunspot numbers using the LSTM-WGAN model}",
      journal = {Frontiers in Astronomy and Space Sciences},
         year = 2025,
        month = feb,
       volume = {12},
          eid = {1541299},
        pages = {1541299},
          doi = {10.3389/fspas.2025.1541299},
       adsurl = {https://ui.adsabs.harvard.edu/abs/2025FrASS..1241299Y}
}

@ARTICLE{2022arXiv220305757D,
       author = {{Dang}, Yuchen and {Chen}, Ziqi and {Li}, Heng and {Shu}, Hai},
        title = "{A comparative study of non-deep learning, deep learning, and ensemble learning methods for sunspot number prediction}",
      journal = {arXiv e-prints},
         year = 2022,
        month = mar,
          eid = {arXiv:2203.05757},
        pages = {arXiv:2203.05757},
          doi = {10.48550/arXiv.2203.05757},
archivePrefix = {arXiv},
       eprint = {2203.05757},
 primaryClass = {astro-ph.SR},
       adsurl = {https://ui.adsabs.harvard.edu/abs/2022arXiv220305757D}
}

@ARTICLE{Martens2012,
       author = {{Martens}, P.~C.~H. and {Attrill}, G.~D.~R. and {Davey}, A.~R. and {Engell}, A. and {Farid}, S. and {Grigis}, P.~C. and {Kasper}, J. and {Korreck}, K. and {Saar}, S.~H. and {Savcheva}, A. and {Su}, Y. and {Testa}, P. and {Wills-Davey}, M. and {Bernasconi}, P.~N. and {Raouafi}, N.-E. and {Delouille}, V.~A. and {Hochedez}, J.~F. and {Cirtain}, J.~W. and {DeForest}, C.~E. and {Angryk}, R.~A. and {De Moortel}, I. and {Wiegelmann}, T. and {Georgoulis}, M.~K. and {McAteer}, R.~T.~J. and {Timmons}, R.~P.},
        title = "{Computer Vision for the Solar Dynamics Observatory (SDO)}",
      journal = {\solphys},
         year = 2012,
        month = jan,
       volume = {275},
       number = {1-2},
        pages = {79-113},
          doi = {10.1007/s11207-010-9697-y},
       adsurl = {https://ui.adsabs.harvard.edu/abs/2012SoPh..275...79M}
}

@ARTICLE{Georgoulis2021,
       author = {{Georgoulis}, Manolis K. and {Bloomfield}, D. Shaun and {Piana}, Michele and {Massone}, Anna Maria and {Soldati}, Marco and {Gallagher}, Peter T. and {Pariat}, Etienne and {Vilmer}, Nicole and {Buchlin}, Eric and {Baudin}, Frederic and {Csillaghy}, Andre and {Sathiapal}, Hanna and {Jackson}, David R. and {Alingery}, Pablo and {Benvenuto}, Federico and {Campi}, Cristina and {Florios}, Konstantinos and {Gontikakis}, Constantinos and {Guennou}, Chloe and {Guerra}, Jordan A. and {Kontogiannis}, Ioannis and {Latorre}, Vittorio and {Murray}, Sophie A. and {Park}, Sung-Hong and {von Stachelski}, Samuelvon and {Torbica}, Aleksandar and {Vischi}, Dario and {Worsfold}, Mark},
        title = "{The flare likelihood and region eruption forecasting (FLARECAST) project: flare forecasting in the big data \& machine learning era}",
      journal = {Journal of Space Weather and Space Climate},
         year = 2021,
        month = may,
       volume = {11},
          eid = {39},
        pages = {39},
          doi = {10.1051/swsc/2021023},
archivePrefix = {arXiv},
       eprint = {2105.05993},
 primaryClass = {astro-ph.SR},
       adsurl = {https://ui.adsabs.harvard.edu/abs/2021JSWSC..11...39G}
}

\end{document}